\documentclass[12pt,preprint]{aastex}
\usepackage{amsmath}

\slugcomment{Accepted for publication in The Astrophysical Journal}

\shorttitle{OH Flares in IRAS$\,$18566+0408}
\shortauthors{Al-Marzouk, et al.}

\def \kms {$\,$km s$^{-1}$}

\def \h{$^{\rm h}$}
\def \m{$^{\rm m}$}
\def\s{\hbox{$^{\rm s}$}}
\def \d {$\arcdeg$}
\def \am {$\arcmin$}
\def \as {\hbox{$\arcsec$}}

\begin{document}

\title{Discovery of 6.035$\,$GHz Hydroxyl Maser Flares in IRAS$\,$18566+0408}

\author{A. A. Al-Marzouk\altaffilmark{1},
E. D. Araya\altaffilmark{1},
P. Hofner\altaffilmark{2,}\altaffilmark{3},
S. Kurtz\altaffilmark{4},
H. Linz\altaffilmark{5}, \&
L. Olmi\altaffilmark{6,}\altaffilmark{7}.}

\altaffiltext{1}{Physics Department, Western Illinois University, 
1 University Circle, Macomb, IL 61455, USA.}
\altaffiltext{2}{New Mexico Institute of Mining and Technology, 
Physics Department, 801 Leroy Place,
Socorro, NM 87801, USA.}
\altaffiltext{3}{National Radio Astronomy Observatory, P.O. 
Box 0, Socorro, NM 87801, USA.}
\altaffiltext{4}{Centro de Radioastronom\'{\i}a y Astrof\'{\i}sica,
Universidad Nacional Aut\'onoma de M\'exico, Apdo. Postal 3-72, 
58090, Morelia, Michoac\'an, Mexico.}
\altaffiltext{5}{Max--Planck--Institut f\"ur Astronomie, K\"onigstuhl 17,
D--69117 Heidelberg, Germany.}
\altaffiltext{6}{INAF, Osservatorio Astrofisico di Arcetri, Largo E. 
Fermi 5, I-50125 Firenze, Italy.}
\altaffiltext{7}{University of Puerto Rico at Rio Piedras, Physics
Department, P.O. Box 23343, San Juan, PR 00931, USA.}

\begin{abstract}

We report the discovery of 6.035$\,$GHz hydroxyl (OH) maser flares 
toward the massive star forming region IRAS$\,$18566+0408 (G37.55+0.20),
which is the only region known to show periodic formaldehyde 
(4.8$\,$GHz H$_{2}$CO) and methanol (6.7$\,$GHz CH$_{3}$OH) 
maser flares. The observations were conducted between October 
2008 and January 2010 with the 305$\,$m Arecibo Telescope in 
Puerto Rico. We detected two flare events, one in March 2009, 
and one in September to November 2009. 
The OH maser flares are not simultaneous with the H$_2$CO flares, 
but may be correlated with CH$_3$OH flares from a component 
at corresponding velocities.
A possible correlated variability of OH and CH$_{3}$OH 
masers in IRAS$\,$18566+0408 is consistent with a common excitation 
mechanism (IR pumping) as predicted by theory.

\end{abstract}


\section{Introduction}

Maser lines from different molecular species,  
including water, hydroxyl, and methanol, are  
common observational phenomena associated with 
massive star forming regions.
The relation between different types of masers found around
young stellar objects may yield important 
information about the evolutionary 
state of regions (e.g., Szymczak \& Gerard 2004; Breen et al. 2010). 
Moreover, excitation of different maser species may
occur within overlapping ranges of physical conditions, 
thus, masers of different species originating from the 
same volume of gas can help narrow the physical conditions
of specific regions (e.g., Edris et al. 2005; see also Fish 2007). 

Masers from hydroxyl (OH) transitions are the prototypical 
example of astrophysical masers; indeed, astrophysical 
masers were first detected in OH (Weaver et al. 1965).
The ground state lines at 1612, 1665, 1667, and
1720$\,$MHz are the most commonly observed OH masers, however, masers
from a number of exited states have also been detected (e.g., 
Baudry \& Desmurs 2002).  
OH masers have been found in a variety of environments, from
galactic star forming regions (e.g., Argon et al. 2000; Fish et al. 2005) 
and supernova remnants (e.g., Brogan et al. 2000), to 
extragalactic environments (e.g., Darling \& Giovanelli 2002; 
Baan et al. 1982). In the 
case of massive star forming regions, many OH masers are
found associated with compact H{$\,$\small II} regions 
(e.g., Fish et al. 2005), however a significant fraction of 
OH masers are also associated with earlier phases of 
massive star formation (e.g., Forster \& Caswell 2000). 

One maser line that is studied in 
massive star forming regions corresponds to the 
F = 3$^-$ $-$ 3$^+$ hyperfine transition of the 
$^2\Pi_{3/2}$ (J = 5/2) excited state of OH
at a frequency of 6.035$\,$GHz (Knowles 
et al. 1976). First detected by Yen et al. (1969)
toward W3(OH), maser lines from this transition 
have been found associated with massive
star formation in galactic and extragalactic 
environments (Caswell \& Vaile 1995;
Caswell 1995). The 6.035$\,$GHz OH line shows
a large flux density range in galactic maser
sources, from more than 100$\,$Jy to $\sim 0.1\,$Jy
(e.g., Baudry et al. 1997). The linewidths are 
narrow; for example, most of the 6.035$\,$GHz OH masers 
in the sample of Baudry et al. (1997) were more narrow
than 0.35$\,$\kms~(some were narrower than 0.18\kms) 
with no correlation between linewidth and peak intensity.
Many 6.035$\,$GHz maser lines show a high fraction of 
circular polarization consistent with Zeeman pairs, allowing for
the measurement of magnetic field strengths (e.g., 
Caswell \& Vaile 1995; Caswell et al. 2009; Fish \& Sjouwerman 2010). 
Most models agree that gas 
densities $\ga 10^7$~cm$^{-3}$ are necessary to enable the population 
inversion (e.g., Baudry et al.~1997, Cragg et al.~2002).

\newpage

The 6.035$\,$GHz OH masers often exhibit variability 
on timescales of months to years.
Nevertheless, some maser sources have shown relatively constant 
spectral profiles, in particular, the OH maser in W3(OH) 
showed almost no variability over two decades 
(Baudry et al. 1997). 

A remarkable characteristic of 6.035$\,$GHz OH
masers is their association with 6.7$\,$GHz
CH$_3$OH masers. For example, Caswell (1997) 
reported interferometric observations of a sample of 
30 massive star forming regions and found 
multiple examples of groups of 
6.035$\,$GHz OH and 6.7$\,$GHz CH$_3$OH
masers coincident within $\sim 1$\arcsec~(see also Etoka et al. 2005). 
The association between these two maser species
is particularly interesting in the context of the discovery
of periodic CH$_3$OH maser flares in
massive star forming regions (Goedhart et al. 2004, 2009;
van der Walt et al. 2009; van der Walt 2011; Szymczak et al. 2011).
Among the periodic maser flare sources known, 
IRAS$\,$18566+0408 (a massive star forming region
at a distance of 6.7$\,$kpc) is unique because it 
exhibits periodic flares of 6.7$\,$GHz CH$_{3}$OH
{\it and} 6$\,$cm H$_{2}$CO masers (Araya et al. 2010). Here
we present the results of monitoring observations of the
6.035$\,$GHz OH maser in IRAS$\,$18566+0408.

\section{Observations and Data Reduction}

The observations 
were conducted with the 305$\,$m Arecibo Telescope\footnote{The Arecibo 
Observatory is operated by SRI International under a cooperative 
agreement with the National Science Foundation (AST-1100968), and in 
alliance with Ana G. M\'endez-Universidad Metropolitana, and 
the Universities Space Research Association.}
in Puerto Rico between
October 2008 and January 2010. We monitored the 
6.035$\,$GHz main excited-state line of hydroxyl (OH; 
$\nu_0 = 6035.0932\,$MHz, $^2\Pi_{3/2}$, J = 5/2, 
F = 3$^-$ $-$ 3$^+$)\footnote{JPL spectra
line catalog (Pickett et al. 1998) accessed through the database
for astronomical spectroscopy (splatalogue.net).} 
toward the young massive stellar object IRAS$\,$18566+0408 (pointing
position, R.A. = 18\h59\m09.98\s, Decl. = 
+04\d12\am15.6\as, J2000) for a total of 24 epochs. 
We used the WAPP spectrometer, two orthogonal 
linear polarization setup, 9-level sampling,
3.125$\,$MHz (155\kms) bandwidth, and 2048 channels per 
polarization, resulting in a final channel separation
of 1.53$\,$kHz (0.076\kms). We observed in 
position-switching (on-off) mode, with 
integration times of 1 to 5 minutes on-source
per run. The reference (off) position was selected
to cover the same hour-angle and declination
as the on-source observations, with angular offsets of 2 to 6 
minutes East from the Right Ascension of the source. 
The center bandpass
LSR velocity was set to 85\kms. Data reduction was done
in IDL using specialized reduction routines
provided by the Arecibo Observatory. After checking
for consistency, we averaged the polarizations and 
subtracted linear baselines. 
The spectra were imported to CLASS\footnote{CLASS is part of the
GILDAS software package developed by IRAM.} 
to measure line parameters and for further analysis.

The cryogenics system of the C-Band High receiver 
of the Arecibo Telescope was not always available, 
thus, the system temperatures ranged from $\sim$30$\,$K
(when the cryogenics were operational) to more than
200$\,$K (when the cryogenics were turned off).
This resulted in rms noise in the spectra of
$\sim 20\,$mJy (with cryogenics) to more than
100$\,$mJy (without cryogenics). 

The calibrator B1857+129 was observed in every run for
pointing and system checking (1$\,$min on-source observations). 
The pointing was better than 15\arcsec~(typically better than 
10\arcsec). We measured a telescope beam size of  
$\sim 44$\arcsec~(at 6.6$\,$GHz), and a
gain of $\sim6\,$K$\,$Jy$^{-1}$. 

We also observed the 6.035$\,$GHz OH maser
source G34.26+0.15 (pointing position, 
R.A. = 18\h53\m18.5\s, 
Decl. = +01\d14\am59\as, J2000; 40\kms~LSR 
central bandpass velocity) in most of the runs
with the same spectral setup as the IRAS$\,$18566+0408
observations. We observed G34.26+0.15 
for system checking; in particular, as a positive
control for detection of OH masers with the
warm C-Band High receiver. We detected several OH masers 
in G34.26+0.15 (see Table~1). 
We also found weak ($\sim 50\,$mJy) 
broad ($FWHM \ga 10\,\text{km s}^{-1}$) OH absorption in G34.26+0.15. 
The absorption line was almost 
undetectable in the unsmoothed spectra, 
but after substantial smoothing (channel width of $\sim$1\kms),
we were able to detect the absorption in most runs. 
After averaging all data and smoothing to a channel 
width of 1.2\kms~(6$\,$mJy rms), the line parameters of the 
absorption line were $S_\nu = -65\,$mJy, $V_{LSR} = 50$\kms, 
$FWHM = 18$\kms. Absorption overlapping with
6.035$\,$GHz maser lines has been detected toward
other massive star forming regions (e.g., Baudry et al. 1997).

The upper panel of Fig.~1. shows a typical spectrum of the 
6.035$\,$GHz OH maser in G34.26+0.15, obtained on 2008 November 18.
None of the maser components detected in G34.26+0.15 showed
clear variability (see Table~1).  As an example, the light curves
of the two brightest components are shown in the lower panel of
Fig.~1.  For both velocity components, $\chi$-squared fits
(with weighting for the uncertainty of each point) show that the
individual data points are consistent within 3$\sigma$ of the 
linear fits. In other words, we did not detect
significant short-term varibility (flares) in any of the G34.26+0.15 
maser components.

\newpage

As part of our monitoring program, we also observed the 
OH transitions at 6.016$\,$GHz, 6.030$\,$GHz, and
6.049$\,$GHz with the same spectral configuration of the
6.035$\,$GHz observations. No lines were detected toward 
IRAS$\,$18566+0408 at the same rms levels of the 6.035$\,$GHz 
data (see Table~2).

\section{Results}

We detected 6.035$\,$GHz OH maser emission in IRAS$\,$18566+0408
in four out of 24 observational epochs. This is the first detection
of 6.035$\,$GHz OH maser emission in IRAS$\,$18566 +0408.
We list in Table~2 
the rms of each run, and the line parameters of the detections. 
The maser at 85.8\kms~was detected at all four epochs. At the first
of the four epochs (2009 March 11), a second maser was detected
at 89.0\kms.
The flux densities of the 85.8\kms~maser in the two 
orthogonally linear polarizations were consistent within 2.6$\,\sigma$
in all runs, whereas the 89.0\kms~maser showed consistent flux
density between the two polarizations within 3.6$\,\sigma$. There could
be some linearly polarized emission at $\la 3\,\sigma$ 
levels; our data are not suitable for a more 
precise determination\footnote{There are examples of 
sources with significant 6.035$\,$GHz OH 
linear polarization (e.g., Knowles et al. 1976).}.
The telescope was not configured to record all
four Stokes parameters; hence, it is not possible to extract information 
about the degree of circular polarization from the data.

Figure~2 shows the spectra obtained on 2009 March 11 (detection of two
lines), 2009 August 15 (non-detection), and 2009 November 07 
(detection of a single line). It is clear from the figure that 
the flux density of the maser varies with time; specifically, we 
detected two flare events. We found no significant change in 
linewidth or peak velocity of the 85.8\kms~maser.

\vspace{-0.5cm}

\section{Discussion}

We show in the upper panel of Fig.~3 the light curve of 
the 6.035$\,$GHz OH maser component at 85.8\kms, including 
the four detections and all 3$\,\sigma$ upper limits. 
We detected two flare events: the first a single epoch detection in 2009
March, and the second a series of three detections from 2009 September
through 2009 November.  Given the null detections before and after each
of these flares obtained when the cryogenics were operational (low rms), 
we can restrict the two events to a maximum duration of approximately
8 and 5 months, respectively.

As mentioned in the introduction, IRAS$\,$18566+0408 is
the only region known to exhibit quasi--periodic 6$\,$cm H$_2$CO and 
6.7$\,$GHz CH$_3$OH flares. 
In the lower three panels of Fig.~3 we show 
the light curves of two 6.7$\,$GHz CH$_{3}$OH maser components
and the light curve of the 6$\,$cm H$_2$CO maser (data from 
Araya et al. 2010, and {\it in prep.}). 
While the 87.8\kms~CH$_3$OH maser component shows very similar 
flares to those of the H$_2$CO maser, the flares of the CH$_3$OH maser 
component at 86.4\kms~are not as well-defined, 
and the flares may have a delay of 1 to 3 
months with respect to H$_2$CO. We note that in addition to 
the two (weak) flare events of the 86.4\kms~CH$_3$OH maser shown in 
Fig.~3, two other flare events of this velocity component have
been detected (Araya et al. 2010). Thus, there is 
evidence that the 86.4\kms~CH$_3$OH maser 
shows quasi-periodic flares, although not as clearly defined
or as regular as the H$_2$CO and 87.8\kms~CH$_3$OH maser lines. 

As seen in Fig.~3, the 85.8\kms~OH maser reported in this work has a similar 
variability behavior to the 86.4\kms~CH$_3$OH maser. 
Thus, the light curves of the two CH$_3$OH masers 
associate the 86.4\kms~maser with the OH, and the 
87.8\kms~maser with the H$_2$CO. The velocity difference 
between the two methanol masers is quite small (1.4\kms)
but they are found at opposite ends of the CH$_3$OH maser
arc imaged by Araya et al. (2010), with a projected separation
of $\sim 6,000\,$AU along the arc.

Given the similar variability profiles and LSR velocities, both 
OH and 86.4\kms~CH$_3$OH masers could originate from the same volume
of gas. Based on the data reported here, we cannot reliably measure
the time delay between the peak of the OH and H$_2$CO flares, but we can
rule out simultaneous flares (see Fig.~3). Our data are consistent
with a delay of 1 to 3 months between the OH and H$_2$CO flares
just as observed between the H$_2$CO and 86.4\kms~CH$_3$OH 
maser. Interferometric observations and a longer monitoring
program are needed to confirm the association between the 
86.4\kms~CH$_3$OH and 85.8\kms~OH masers. 

High angular resolution observations have shown an association 
between 6.035$\,$GHz OH and 6.7$\,$GHz CH$_3$OH masers. For example, 
Caswell (1997) found that both maser species often show emission at
similar velocities and co-exist in elongated
structures with projected sizes of $\sim $2,000 to 6,000$\,$AU.
The discovery of 6.035$\,$GHz OH flares and possible
correlated variability with 6.7$\,$GHz masers brings a new
(time-dependent) aspect to the relation between these maser species. 

The physical mechanism causing the periodic flares of
CH$_3$OH masers detected in a number of sources (e.g., Goedhart et al.
2004) is still unclear. However, van der Walt (2011; see also 
van der Walt et al. 2009) reproduced remarkably well the flare 
profiles observed toward G9.62+0.20E with a colliding wind 
binary (CWB) model, in which the flares are caused by a change in 
the background radio continuum modulated by the orbital 
parameters of a young massive binary. Based only on
the detection of 6.035$\,$GHz OH flares reported here, we 
cannot address whether the CWB model is applicable in the 
case of IRAS$\,$18566+0408. Nevertheless, as discussed by 
Araya et al. (2010), the H$_2$CO and CH$_3$OH maser flares in 
IRAS$\,$18566+0408 are likely caused by a change in the maser 
gains and not by a change in the background continuum. In this 
scenario, the maser gains are modulated by some periodic 
phenomenon external to the maser regions (possibilities 
include periodic accretion events onto a central protobinary system).
If the CH$_3$OH flares are caused by a change in the maser
gain, then correlated variability
with OH masers would indicate a similar excitation mechanism 
for 6.035$\,$GHz OH and 6.7$\,$GHz CH$_3$OH masers. Indeed,
theoretical models have shown that the excitation mechanism 
of class II CH$_3$OH masers is infrared pumping (e.g., Cragg et al. 2005), 
and that the population inversion of 6.035$\,$GHz OH masers
is also predominantly due to infrared radiation (Gray 2001; 
see also Pihlstr\"om et al. 2008, Baudry et al. 1997)\footnote{However, 
collisional excitation may also have a prominent role as indicated
by Cragg et al. (2002).}.

The possibility that the maser flares are caused by gain variability
due to changes in the infrared radiation field can qualitatively
explain some of the differences between the various light-curves.
For example, a 70-day delay of the OH flare following the H$_2$CO flare
could result from the time required for pumping photons to propagate
between the H$_2$CO and the OH maser regions.  A 70 light-day distance
corresponds to 12,000$\,$AU, which is of the same order as the 6,000$\,$AU
projected size of the CH$_3$OH maser arc reported by Araya et al. (2010).
 
If the 87.8\kms~CH$_3$OH and the H$_2$CO
maser regions are closer to the central source of infrared
field variability, then an exponential amplification of a change
in the infrared pumping rate would result in a clear
flare signature. On the contrary, maser regions located at greater distances 
would show a less clear flare signature due to geometrical dilution
of the variable radiation field, optical depth effects, 
and a greater relative contribution
of other sources of pumping photons. It is worth
mentioning that the models of Cragg et al.~(2002) predict 
that the 6035$\,$MHz OH masers appear in zones of high density and high OH 
column density, but relatively low gas temperature. In fact, at 
kinematic temperatures $>70\,$K, the line would eventually be quenched. 
This suggests that H$_2$CO masers
may occur in warmer gas closer to the central energy source than the excited 
OH masers. However, the exact circumstances will become clear only after 
interferometric mapping of the OH maser and the dense molecular 
gas is conducted. 

Interferometric observations are also required to investigate
the relation between the masers discussed here and
ground state OH emission in the region, which has been detected
with single-dish telescopes (Szymczak \& G\'erard 2004). 
For example, Edris et al. (2007) mapped the 1665 and 1667$\,$MHz
masers in this source using the NRAO Green Bank Telescope (GBT;
8\arcmin~beam size).
Although the LSR velocities of the OH ground state masers are 
similar to the CH$_3$OH, H$_2$CO, and 6.035$\,$GHz OH masers, 
the positions of the ground state OH masers obtained with the GBT
do not correspond to the H$_2$CO maser position within the 
quoted errors (Edris et al. 2007).

\section{Summary}

Using the 305$\,$m Arecibo Telescope in Puerto Rico, we 
detected two flare events of the 6.035$\,$GHz OH maser 
toward the massive star forming region IRAS$\,$18566+0408.
This region is the only known source of periodic H$_2$CO
and CH$_3$OH maser flares. Despite poor
sampling of the OH light curve during the flares, our 
observations clearly show that the peaks of the OH flares
were not simultaneous with the H$_2$CO peaks, but rather
had delays of approximately a month or more. In contrast, the peaks of the 
OH flares appear to be correlated with a 6.7$\,$GHz 
CH$_3$OH maser at corresponding LSR velocity.
Our results strengthen the association between
6.035$\,$GHz OH and 6.7$\,$GHz CH$_3$OH masers found 
in previous observational work and are consistent with a similar inversion 
mechanism of these maser species (radiative excitation). The 
delay between the H$_2$CO and OH flares might be caused by 
the difference in arrival times of pumping photons 
between the two maser regions. Consequently, interferometric 
observations will be the natural next step in order to pinpoint the 
exact location of the OH masers. A more extended monitoring program 
is also needed to confirm the association between the OH and CH$_3$OH masers.

\acknowledgments

We thank an anonymous referee for comments that significantly improved this 
manuscript. E.D.A. acknowledges partial support from the WIU Foundation 
and the Office of Sponsored Projects (faculty proposal planning
stipend, and summer stipend).
P.H. acknowledges partial support from NSF grant AST-0908901. 
S.K. acknowledges support from DGAPA grant IN-101310, UNAM.

\newpage

\begin{figure}
\includegraphics{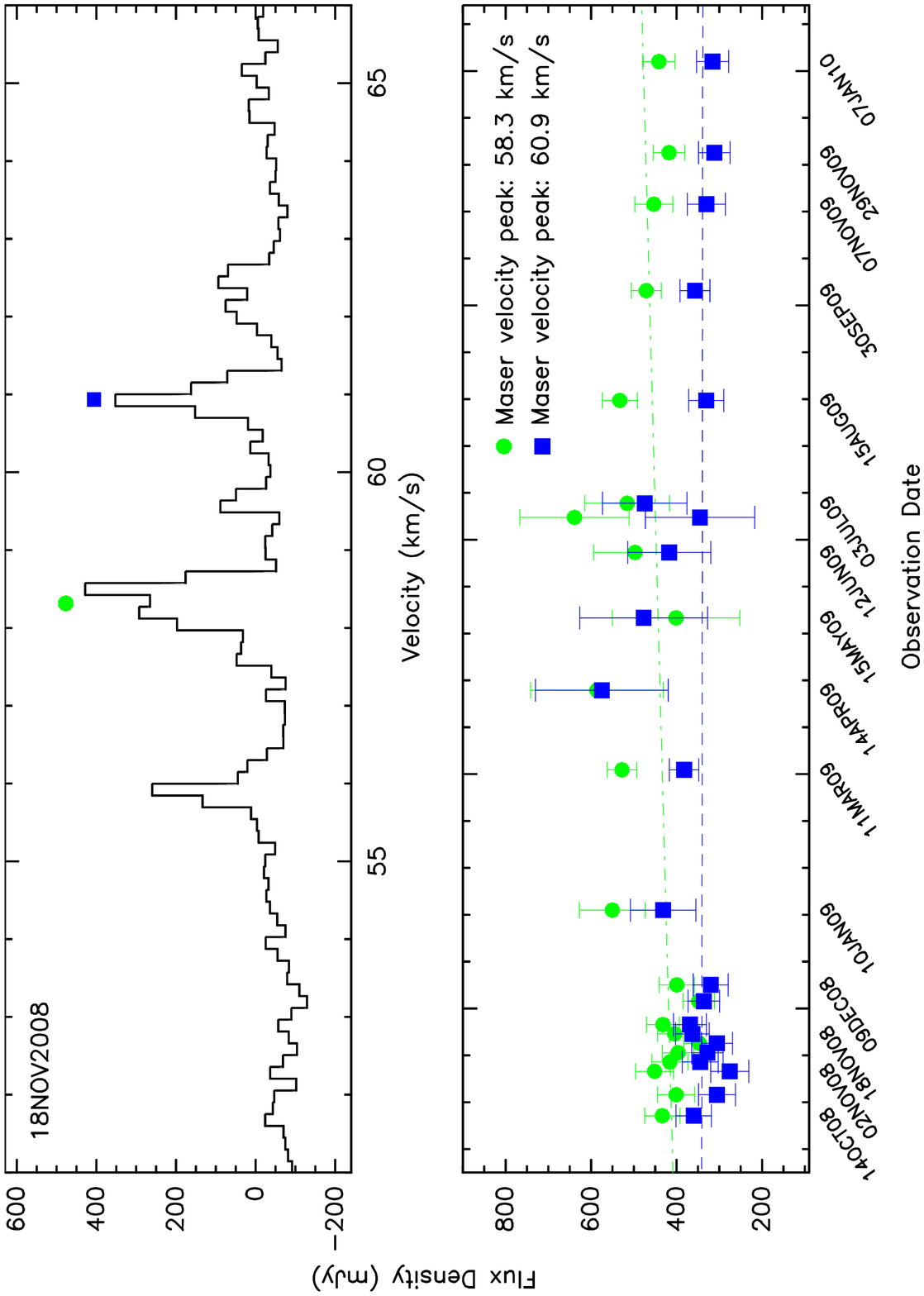}
\vspace*{15.5cm}\caption{System checking observations of
G34.26+0.15. The upper panel shows the 6.035$\,$GHz OH spectrum
from the 2008 November 18 observations (0.15\kms~channel width). 
As an example, the 
light curves of the two brightest maser components are shown 
in the lower panel. The dashed lines are linear fits of the
components in the form $S_\nu(t) = a\,(t_{JD}-2454754.40) + b$, where 
$a = 0.15 \pm 0.07\,$mJy$\,$day$^{-1}$ and $b = 413 \pm 16\,$mJy
for the 58.3\kms~line, and $a = 0.00 \pm 0.05\,$mJy$\,$day$^{-1}$ 
and $b = 341 \pm 11\,$mJy for the 60.9\kms~line.
The length of the error bars is three times 
the rms noise of the respective spectrum. Data points with large 
error bars correspond to observations conducted with no cryogenics.}
\label{f1}
\end{figure}

\newpage

\begin{figure}
\includegraphics{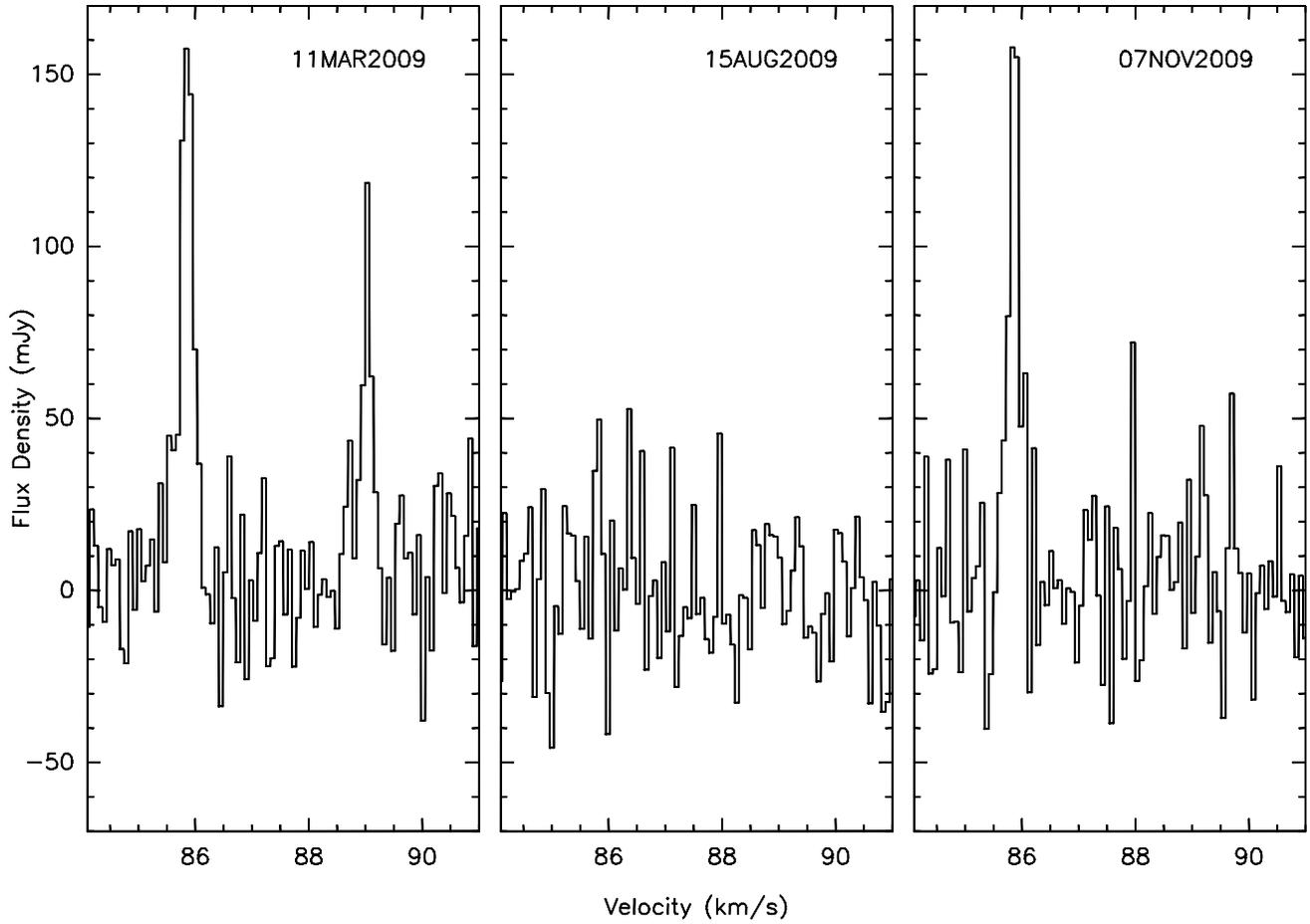}
\vspace*{15.5cm}\caption{Detection of 6.035$\,$GHz OH maser flares 
in IRAS$\,$18566+0408. The first panel shows the two maser components 
detected on 2009 March 11. No maser was detected on
2009 August 15 (middle panel), and a single maser was detected on
2009 November 07. The cryogenics were operational in all three 
epochs shown in the figure.}
\label{f2}
\end{figure}

\newpage

\begin{figure}
\includegraphics{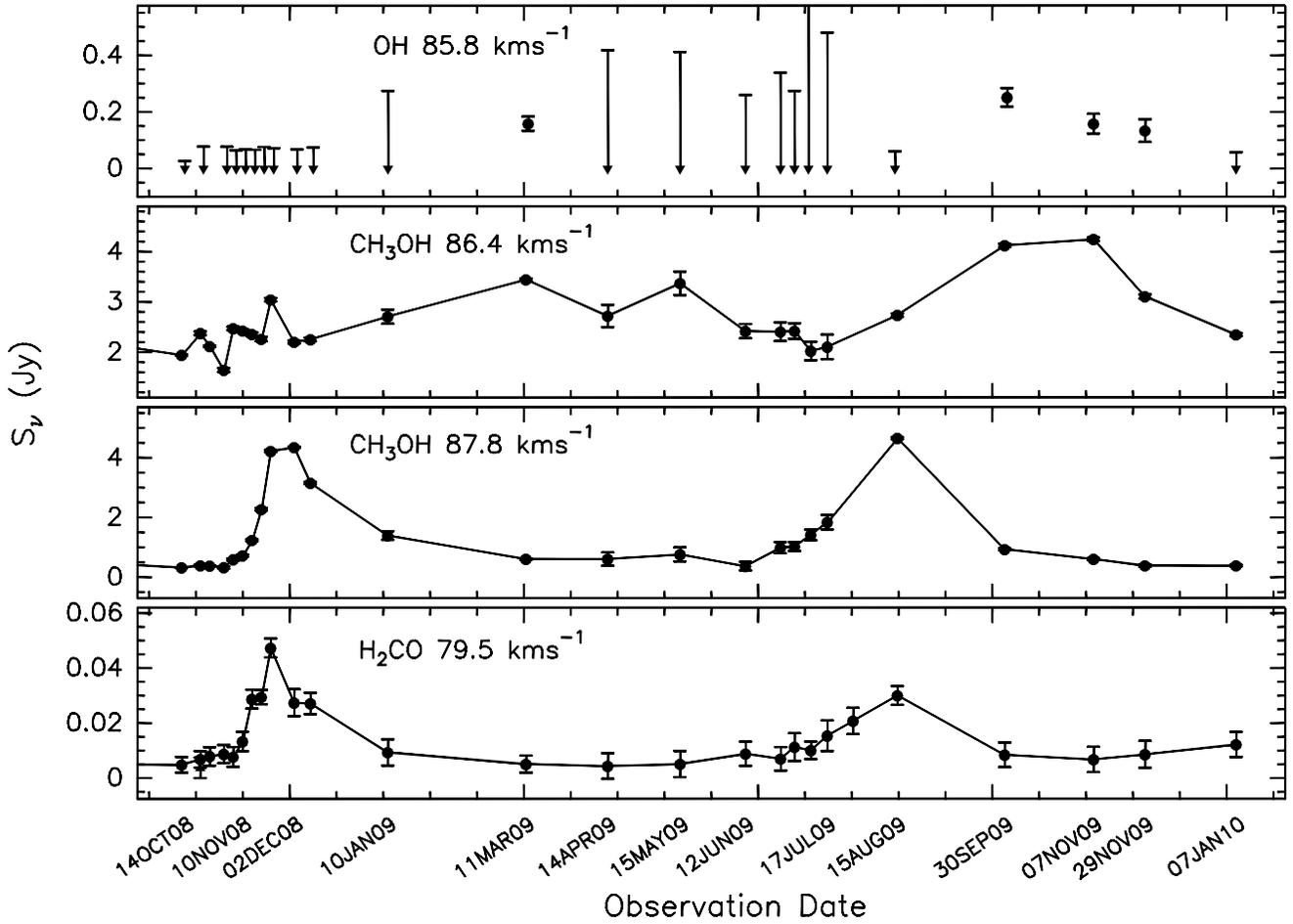}
\vspace*{15.5cm}\caption{Light curve of 6.035$\,$GHz hydroxyl, 
6.7$\,$GHz methanol and 4.8 GHz formaldehyde masers in 
IRAS$\,$18566+0408. The error bars are three times the rms 
noise; three times the rms is also shown as upper limits for the non-detections
(triangles). The upper limit of the 2009 July 09
observations (which is outside of the axis range) is 0.8$\,$Jy. 
The upper panel shows the light curve of the 
OH maser component at the velocity 85.8$\,$\kms. 
The bottom panels show two 6.7$\,$GHz CH$_3$OH
and the 6$\,$cm H$_2$CO maser light curves from 
Araya et al. (2010).}
\label{f3}
\end{figure}

\newpage

\begin{deluxetable}{lccccc}
\tabletypesize{\scriptsize}
\tablecaption{Line Parameters of 6.035$\,$GHz OH in G34.26+0.15\label{tbl-1}}
\tablewidth{0pt}
\tablehead{
\colhead{Date} & \colhead{rms} & \colhead{S$_{\nu}^*$} 
& \colhead{V$_{LSR}$} & \colhead{Width} & \colhead{$\int S_\nu dv$}\\
\colhead{ } & \colhead{(mJy)} & \colhead{(mJy)} & 
\colhead{(\kms) } & \colhead{(\kms)} & \colhead{(mJy\kms)} 
}
\startdata
14OCT2008     & 28    & 290 & 55.83 (0.02) & 0.34 (0.05) & 106  (11)  \\
              &       & 434 & 58.34 (0.02) & 0.63 (0.04) & 289  (14)  \\ 
              &       & 161 & 59.57 (0.03) & 0.32 (0.07) & 55   (10)  \\
              &       & 361 & 60.93 (0.02) & 0.42 (0.04) & 159  (12)  \\
              &       & 136 & 62.25 (0.06) & 0.6  (0.1)  & 90   (13)  \\
23OCT2008     & 29    & 247 & 55.84 (0.02) & 0.37 (0.08) & 98   (13)  \\
              &       & 402 & 58.32 (0.02) & 0.67 (0.04) & 286  (14)  \\
              &       & 156 & 59.61 (0.03) & 0.28 (0.07) & 47   (9)   \\ 
              &       & 306 & 60.95 (0.02) & 0.50 (0.05) & 162  (13)  \\
02NOV2008     & 30    & 291 & 55.86 (0.02) & 0.39 (0.06) & 119  (13)  \\
              &       & 452 & 58.34 (0.02) & 0.65 (0.04) & 310  (15)  \\
              &       & 121 & 59.57 (0.04) & 0.27 (0.07) & 35   (9)   \\
              &       & 277 & 60.95 (0.02) & 0.49 (0.06) & 144  (13)  \\
              &       & 133 & 62.34 (0.06) & 0.6  (0.1)  & 87   (14)  \\
06NOV2008     & 28    & 218 & 55.85 (0.03) & 0.52 (0.07) & 120  (13)  \\
              &       & 417 & 58.34 (0.02) & 0.61 (0.03) & 272  (13)  \\
              &       & 167 & 59.59 (0.03) & 0.25 (0.05) & 44   (9)   \\
              &       & 346 & 60.95 (0.02) & 0.38 (0.04) & 138  (11)  \\
              &       & 124 & 62.29 (0.05) & 0.55 (0.07) & 73   (12)  \\
10NOV2008     & 25    & 242 & 55.84 (0.02) & 0.38 (0.04) & 99   (9)   \\
              &       & 397 & 58.35 (0.01) & 0.64 (0.03) & 270  (12)  \\
              &       & 212 & 59.60 (0.02) & 0.23 (0.04) & 53   (7)   \\
              &       & 329 & 60.93 (0.01) & 0.37 (0.04) & 131  (10)  \\
              &       & 124 & 62.30 (0.04) & 0.54 (0.07) & 71   (10)  \\
14NOV2008     & 24    & 224 & 55.85 (0.02) & 0.44 (0.06) & 105  (11)  \\
              &       & 348 & 58.37 (0.01) & 0.64 (0.04) & 236  (12)  \\
              &       & 146 & 59.60 (0.03) & 0.43 (0.08) & 67   (10)  \\
              &       & 307 & 60.93 (0.01) & 0.40 (0.04) & 132  (10)  \\
18NOV2008     & 26    & 294 & 55.89 (0.01) & 0.35 (0.06) & 108  (12)  \\
              &       & 405 & 58.35 (0.02) & 0.64 (0.04) & 274  (13)  \\
              &       & 146 & 59.63 (0.07) & 0.2  (0.1)  & 38   (20)   \\
              &       & 364 & 60.93 (0.01) & 0.36 (0.03) & 139  (10)  \\
              &       & 110 & 62.30 (0.06) & 0.65 (0.09) & 76   (12)  \\
22NOV2008     & 26    & 240 & 55.87 (0.02) & 0.42 (0.05) & 106  (10)  \\
              &       & 433 & 58.35 (0.01) & 0.65 (0.03) & 299  (12)  \\
              &       & 170 & 59.63 (0.03) & 0.35 (0.07) & 63   (10)  \\
              &       & 370 & 60.93 (0.01) & 0.36 (0.03) & 143  (10)  \\
              &       & 143 & 62.4  (0.1)  & 0.6  (0.3)  & 88   (40)  \\
02DEC2008     & 25    & 241 & 55.81 (0.02) & 0.50 (0.05) & 129  (11)  \\
              &       & 349 & 58.35 (0.02) & 0.72 (0.04) & 267  (12)  \\
              &       & 337 & 60.93 (0.01) & 0.36 (0.03) & 127  (9)   \\
              &       & 100 & 62.27 (0.06) & 0.8  (0.1)  & 83   (12)  \\
09DEC2008     & 27    & 291 & 55.85 (0.02) & 0.34 (0.04) & 105  (11)  \\
              &       & 401 & 58.37 (0.02) & 0.67 (0.04) & 284  (14)  \\
              &       & 160 & 59.57 (0.03) & 0.35 (0.07) & 60   (10)  \\
              &       & 321 & 60.95 (0.02) & 0.38 (0.04) & 129  (11)  \\
              &       & 154 & 62.45 (0.03) & 0.21 (0.05) & 34   (9)   \\
10JAN2009     & 51    & 342 & 55.84 (0.02) & 0.23 (0.09) & 83   (20)  \\
              &       & 551 & 58.37 (0.02) & 0.58 (0.06) & 340  (28)  \\
              &       & 361 & 59.60 (0.03) & 0.17 (0.05) & 67   (15)  \\
              &       & 432 & 60.93 (0.03) & 0.41 (0.08) & 189  (27)  \\
              &       & 289 & 62.41 (0.03) & 0.25 (0.07) & 77   (18)  \\
11MAR2009     & 23    & 329 & 55.9  (0.2)  & 0.35 (0.04) & 124  (11)  \\
              &       & 529 & 58.4  (0.2)  & 0.70 (0.03) & 403  (15)  \\
              &       & 188 & 59.6  (0.2)  & 0.29 (0.05) & 62   (10)  \\
              &       & 383 & 61.0  (0.2)  & 0.48 (0.04) & 188  (13)  \\
              &       & 162 & 62.3  (0.5)  & 0.60 (0.08) & 130  (18)  \\
14APR2009     & 104   & 587 & 58.28 (0.05) & 0.69 (0.09) & 430  (55)  \\
              &       & 576 & 60.93 (0.03) & 0.4  (0.1)  & 259  (48)  \\
15MAY2009     & 100   & 403 & 55.86 (0.06) & 0.7  (0.1)  & 293  (52)  \\
              &       & 402 & 58.25 (0.07) & 0.9  (0.2)  & 375  (63)  \\
              &       & 478 & 60.88 (0.04) & 0.3  (0.1)  & 162  (39)  \\
12JUN2009     & 65    & 310 & 55.84 (0.04) & 0.35 (0.09) & 114  (24)  \\
              &       & 498 & 58.34 (0.03) & 0.66 (0.06) & 348  (30)  \\
              &       & 298 & 59.59 (0.04) & 0.3  (0.1)  & 109  (24)  \\
              &       & 418 & 60.97 (0.02) & 0.31 (0.05) & 138  (21)  \\
27JUN2009     & 85    & 450 & 55.85 (0.02) & 0.24 (0.09) & 116  (26)  \\
              &       & 640 & 58.30 (0.03) & 0.62 (0.06) & 421  (37)  \\
              &       & 346 & 61.00 (0.05) & 0.5  (0.1)  & 184  (35)  \\
03JUL2009     & 66    & 361 & 55.83 (0.06) & 0.5  (0.3)  & 175  (74) \\
              &       & 517 & 58.29 (0.04) & 0.81 (0.09) & 443  (43)  \\
              &       & 475 & 60.98 (0.03) & 0.33 (0.07) & 167  (29)  \\
09JUL2009     & 282   & \nodata & \nodata  & \nodata     & \nodata    \\
15AUG2009     & 28    & 306 & 55.85 (0.02) & 0.56 (0.05) & 184  (13)  \\
              &       & 534 & 58.33 (0.01) & 0.68 (0.03) & 389  (13)  \\
              &       & 200 & 59.59 (0.02) & 0.25 (0.04) & 54   (8)   \\
              &       & 331 & 60.96 (0.01) & 0.43 (0.04) & 152  (11)  \\
              &       & 132 & 62.26 (0.05) & 0.67 (0.09) & 95   (13)  \\
30SEP2009     & 23    & 228 & 55.84 (0.02) & 0.55 (0.06) & 132  (12)  \\
              &       & 472 & 58.31 (0.01) & 0.71 (0.03) & 355  (13)  \\
              &       & 208 & 59.59 (0.02) & 0.33 (0.05) & 74   (9)   \\
              &       & 358 & 60.94 (0.01) & 0.41 (0.03) & 158  (10)  \\
              &       & 123 & 62.25 (0.05) & 0.64 (0.08) & 84   (11)  \\
07NOV2009     & 30    & 280 & 55.88 (0.02) & 0.40 (0.04) & 118  (11)  \\
              &       & 454 & 58.33 (0.02) & 0.67 (0.03) & 323  (10)  \\
              &       & 185 & 59.59 (0.03) & 0.35 (0.06) & 68   (10)  \\
              &       & 331 & 60.95 (0.02) & 0.45 (0.04) & 157  (12)  \\
              &       & 132 & 62.27 (0.03) & 0.56 (0.03) & 78   (11)  \\
29NOV2009     & 25    & 302 & 55.83 (0.01) & 0.32 (0.03) & 103  (9)   \\
              &       & 419 & 58.35 (0.01) & 0.64 (0.03) & 286  (11)  \\
              &       & 208 & 59.62 (0.02) & 0.21 (0.04) & 46   (7)   \\
              &       & 313 & 60.94 (0.01) & 0.35 (0.03) & 117  (9)   \\
              &       & 109 & 62.34 (0.05) & 0.58 (0.09) & 67   (11)  \\
07JAN2010     & 25    & 291 & 55.83 (0.02) & 0.40 (0.04) & 125  (10)  \\
              &       & 442 & 58.35 (0.01) & 0.66 (0.03) & 309  (12)  \\
              &       & 171 & 59.57 (0.02) & 0.30 (0.05) & 55   (8)   \\
              &       & 317 & 60.93 (0.02) & 0.40 (0.04) & 136  (11)  \\
              &       & 102 & 62.16 (0.07) & 0.8  (0.1)  & 90   (13)  \\
\enddata
\tablenotetext{~}{~Line parameters and rms noise were obtained 
after smoothing the spectra to a final channel width of 0.15\kms.
We include in parenthesis 1$\,\sigma$ statistical errors from the fit. 
Only spectral lines brighter than 4$\,\sigma$ are listed.
The OH components at 58.3 and 62.3\kms~are
superposition of narrower lines, however single Gaussian
profiles gave acceptable fits given the channel width and
signal-to-noise.}
\tablenotetext{(*)}{~After smoothing the data to a channel width of 
$\sim 1$\kms, we found evidence in most of the scans for
a broad, weak OH absorption line at $\sim 50$\kms~LSR peak velocity. 
After averaging all the data and 
smoothing the spectrum to a channel width of 1.2\kms~(6$\,$mJy rms), 
the line parameters of the absorption line were: 
$S_\nu = -65\,$mJy, $V_{LSR} = 50$\kms, 
$FWHM = 18$\kms.}
\end{deluxetable}

\newpage

\begin{deluxetable}{lcc ccc}
\tabletypesize{\scriptsize}
\tablecaption{Line Parameters of the 6.035$\,$GHz OH Masers in 
IRAS$\,$18566+0408\label{tbl-1}}
\tablewidth{0pt}
\tablehead{
\colhead{Observation} & \colhead{rms} & \colhead{S$_{\nu}$} 
& \colhead{V$_{LSR}$} & \colhead{Width} & \colhead{$\int S_\nu dv$}\\
\colhead{Date} & \colhead{(mJy)} & \colhead{(mJy)} & 
\colhead{(\kms) } & \colhead{(\kms)} & \colhead{(mJy\kms)} 
}
\startdata
14OCT2008     & 9.0& \nodata & \nodata & \nodata  & \nodata  \\
23OCT2008     & 26 & \nodata & \nodata & \nodata  & \nodata   \\
02NOV2008     & 25 & \nodata & \nodata & \nodata  & \nodata   \\
06NOV2008     & 21 & \nodata & \nodata & \nodata  & \nodata   \\
10NOV2008     & 22 & \nodata & \nodata & \nodata  & \nodata   \\
14NOV2008     & 22 & \nodata & \nodata & \nodata  & \nodata   \\
18NOV2008     & 25 & \nodata & \nodata & \nodata  & \nodata   \\
22NOV2008     & 24 & \nodata & \nodata & \nodata  & \nodata   \\
02DEC2008     & 22 & \nodata & \nodata & \nodata  & \nodata   \\
09DEC2008     & 25 & \nodata & \nodata & \nodata  & \nodata   \\
10JAN2009     & 91 & \nodata & \nodata & \nodata  & \nodata   \\
11MAR2009     & 17 & 158 & 85.85 (0.01) & 0.29 (0.04) & 48.7 (4.8)   \\
              &    & 106 & 89.03 (0.02) & 0.20 (0.06) & 23.0 (5.2)   \\
14APR2009     & 139 & \nodata & \nodata & \nodata  & \nodata   \\
15MAY2009     & 137 & \nodata & \nodata & \nodata  & \nodata   \\
12JUN2009     & 86  & \nodata & \nodata & \nodata  & \nodata  \\
27JUN2009     & 113 & \nodata & \nodata & \nodata  & \nodata   \\
03JUL2009     & 91  & \nodata & \nodata & \nodata  & \nodata   \\
09JUL2009     & 261 & \nodata & \nodata & \nodata  & \nodata   \\
16JUL2009     & 160 & \nodata & \nodata & \nodata  & \nodata   \\
15AUG2009     & 20  & \nodata & \nodata & \nodata  & \nodata \\
30SEP2009     & 22  & 251 & 85.82 (0.01) & 0.30 (0.02) & 80.7 (5.0)   \\
07NOV2009     & 23  & 158 & 85.86 (0.01) & 0.24 (0.04) & 40.9 (5.0)   \\
29NOV2009     & 27  & 134 & 85.82 (0.02) & 0.23 (0.05) & 32.2 (5.4)   \\
07JAN2010     & 19  & \nodata & \nodata & \nodata  & \nodata   \\
~\\
\enddata
\tablenotetext{~}{1$\,\sigma$ statistical errors from the fit are 
shown in parentheses.}
\end{deluxetable}


\begin{thebibliography}{}


\bibitem[Araya et al. (2010)]{araya10} Araya, E. D., Hofner, P., Goss, W. M., Kurtz, S., Richards, A. M. S., Linz, H., Olmi, L., \& Sewi{\l}o, M. 2010, ApJL, 717, 133

\bibitem[Argon et al. (2000)]{argon00} Argon, A. L., Reid, M. J., \& Menten, K. M. 2000, ApJS, 129, 159

\bibitem[Baan et al. (1982)]{baan82} Baan, W. A., Wood, P. A. D., \& Haschick, A. D. 1982, ApJL, 260, 49

\bibitem[Baudry et al. (1997)]{Baudry1997} Baudry, A., Desmurs, J. F., Wilson, T. L. \& Cohen, R. J. 1997, A\&A, 325, 255

\bibitem[Baudry \& Desmurs (2002)]{baudry02} Baudry, A. \& Desmurs, J. F. 2002, A\&A, 394, 107

\bibitem[Breen et al. (2010)]{breen10} Breen, S. L., Caswell, J. L., Ellingsen, S. P., \& Phillips, C. J. 2010, MNRAS, 406, 1487

\bibitem[Brogan et al. (2000)]{brogan00} Brogan, C. L., Frail, D. A., Goss, W. M., Troland, T. H. 2000, ApJ, 537, 875

\bibitem[Caswell, J. L. (1995)]{caswell1995} Caswell, J. L. 1995, MNRAS, 272, 31L

\bibitem[Caswell, J. L. (1997)]{caswell1997} Caswell, J. L. 1997, MNRAS, 289, 203

\bibitem[Caswell, et al. (2009)]{caswell2009} Caswell, J. L., Kramer, B. Hutawarakorn, \& Reynolds, J. E. 2009, MNRAS, 398, 528

\bibitem[Caswell \& Vaile (1995)]{caswellvaile1995} Caswell, J. L., \& Vaile, R. A. 1995, MNRAS, 273, 328

\bibitem[Cragg et al. (2002)]{cragg02} Cragg, D. M., Sobolev, A. M., \& Godfrey, P. D. 2002, MNRAS, 331, 521

\bibitem[Cragg et al. (2005)]{cragg05} Cragg, D. M., Sobolev, A. M., \& Godfrey, P. D. 2005, MNRAS, 360, 533 

\bibitem[Darling (2002)]{darling02} Darling, J. \& Giovanelli, R. 2002, AJ, 124, 100

\bibitem[Edris et al. (2005)]{edris2005} Edris, K. A., Fuller, G. A., Cohen, R. J. \& Etoka, S. 2005, A\&A, 434, 213

\bibitem[Edris et al. (2007)]{edris2007} Edris, K. A., Fuller, G. A., \& Cohen, R. J. 2007, A\&A, 465, 865


\bibitem[Etoka et al. (2005)]{etoka05} Etoka, S., Cohen, R. J., \& Gray, M. D. 2005, MNRAS, 360, 1162

\bibitem[Fish (2007)]{fish07} Fish, V. L. 2007, in IAU Symp. 242, Astrophysical Masers and Their Environments, ed. J. Chapman, \& W. A. Baan (Cambridge: Cambridge Univ. Press), 71

\bibitem[Fish et al. (2005)]{fish05} Fish, V. L., Reid, M. J., Argon, A. L., Zheng, X.-W. 2005, ApJS, 160, 220

\bibitem[Fish \& Sjouwerman (2010)]{fish10} Fish, V. L., \& Sjouwerman, L. O. 2010, ApJ, 716, 106


\bibitem[Forster \& Caswell (2000)]{forster00} Forster, J. R. \& Caswell, J. L. 2000, ApJ, 530, 371

\bibitem[Goedhart et al. (2004)]{goed04} Goedhart, S., Gaylard, M. J., \& van der Valt, D. J. 2004, MNRAS, 355, 553

\bibitem[Goedhart et al. (2009)]{goed09} Goedhart, S., Langa, M. C.,Gaylard, M. J., \& van der Walt, D. J. 2009, MNRAS, 398, 995

\bibitem[Gray (2001)]{gray01} Gray, M. D. 2001, MNRAS, 324, 57

\bibitem[Knowles et al. (1976)]{knowles76} Knowles, S. H., Caswell, F. L., Goss, W. M. 1976, 175, 537

\bibitem[Pickett et al. (1998)]{pickett98} Pickett, H. M., Poynter, R. L., Cohen, E. A., Delitsky, M. L., Pearson, J. C., \& Muller, H. S. P. 1998, J. Quant. Spectrosc. Radiat. Transfer, 60, 883

\bibitem[Pihlstrom et al. (2008)]{pihlstrom08} Pihlstr\"om, Y. M., Fish, V. L., Sjouwerman, L. O., Zschaechner, L. K., Lockett, P. B., \& Elitzur, M. 2008, ApJ, 676, 371

\bibitem[Szymczak \& Gerard (2004)]{szymczak04} Szymczak, M., \& G\'erard, E. 2004, A\&A, 414, 235

\bibitem[Szymczak et al. (2011)]{szymczak11} Szymczak, M., Wolak, P., Bartkiewicz, A., \& van Langevelde, H.J. 2011, A\&A, 531, 3L

\bibitem[van der Walt (2011)]{vdw11} van der Walt, D. J. 2011, AJ, 141, 152

\bibitem[van der Walt et al. (2009)]{vdw09} van der Walt, D. J., Goedhart, S., \& Gaylard, M. J. 2009, MNRAS, 398, 961

\bibitem[Weaver et al. (1965)]{weaver65} Weaver, H., Williams, D. R. W., Dieter, N. H., \& Lum, W. T. 1965, Nature, 208, 29

\bibitem[Yen et al. (1969)]{yen69} Yen, J. L., et al. 1969, ApJL, 156, 27


\end{thebibliography}
\end{document}